\documentclass[aps,twocolumn,a4paper,showkeys,showpacs,prb]{revtex4-1}

\usepackage{graphicx}
\usepackage{dcolumn}
\usepackage{bm}
\usepackage{amsmath, amssymb, graphics, setspace}
\usepackage{hyperref}
\usepackage{color}

\newcommand{\mathsym}[1]{{}}
\newcommand{\unicode}[1]{{}}
\newcommand{\be}{\begin{eqnarray}}
\newcommand{\ee}{\end{eqnarray}}

\begin{document}

\title{Possibility of chiral $d$-wave state in the hexagonal pnictide superconductor SrPtAs}

\author{Hikaru Ueki, Ryota Tamura, and Jun Goryo}
\affiliation{Department of Mathematics and Physics, Hirosaki University, 036-8561 Hirosaki, Japan }

\date{\today}

\begin{abstract}

We discuss the type of pairing in the hexagonal pnictide superconductor SrPtAs, taking into account its multiband structure. The topological chiral $d$-wave state with time-reversal-symmetry breaking has been anticipated from the spontaneous magnetization observed by the muon-spin-relaxation experiment.  We point out in this paper that the recent experimental reports on the nuclear-spin-lattice relaxation rate $T_1^{-1}$ and superfluid density $n_s(T)$, which seemingly support the conventional $s$-wave pairing, are also consistent with the chiral $d$-wave state. The compatibility of the gap and multiband structures is crucial in this argument. 
We propose that the measurement of the bulk quasiparticle density of states would be useful for the distinction between two pairing states.

\end{abstract}

                              
\maketitle

\section{Introduction}
 
The first hexagonal pnictide superconductor SrPtAs \cite{Nishikubo-Kudo-Nohara} ($T_c=2.4 K$) has received attention, since the muon-spin-relaxation ($\mu$SR) experiment \cite{Biswas-etal} 
observes the internal magnetization below $T_c$. The result suggests the spontaneous time-reversal symmetry (TRS) breaking in
the superconducting state. From the group theoretical consideration \cite{Goryo-Fischer-Sigrist,Fischer-Goryo} and functional renormalization group (FRG) analysis, \cite{Fischer-etal} 
the most probable pairing symmetry is the topological chiral $d$-wave ($d_{x^2-y^2} \pm i d_{xy}$-wave) state with TRS breaking. 
This state has non-zero Chern number\cite{TKNN} and supports the surface bound states with chiral energy spectrum.\cite{Volovik,Fischer-etal} 
Especially in SrPtAs, it is expected that the chiral surface state causes spontaneous spin current and spin polarization,\cite{Goryo-etal-spin} the origin of which is the staggered anti-symmetric spin-orbit coupling (SOC) coming from
the hexagonal bi-layer structure of the crystal with local lack of inversion symmetry.\cite{Sigrist-review} 

We may explain intuitively the stability of the chiral $d$-wave pairing in SrPtAs. The hexagonal structure of the crystal plays a role for supporting the chiral $d$-wave state, since 
there is the two-dimensional (2D) irreducible representation with $d_{x^2-y^2}$- and $ d_{xy}$-wave functions in the crystal symmetry, 
and then the chiral $d$-wave pairing is easily obtained as the mixing of these two basis with relative phase $\pm \pi/2$. \cite{Goryo-Fischer-Sigrist,Fischer-Goryo} 
Moreover, the band structure also assists the condensation energy gain, 
since quasi-2D bands with fully-gapped quasiparticle excitations dominate significantly compared to 
a minor three-dimensional (3D) band with point-nodal excitation.\cite{Fischer-etal,Youn-etal,Youn-etal-2} 

On the other hand, there are still some controversies on the chiral $d$-wave pairing. The nuclear spin-lattice relaxation rate $T_1^{-1}$ measured by the nuclear quadrupole resonance 
shows the Hebel-Slichter (HS) peak near $T_c$ and exponential decay in the low temperature region.\cite{Matano-etal} 
It has also been found from the magnetic-penetration-depth measurement that superfluid density $n_s(T)$ exhibits the Arrhenius-type behavior (i.e., approaches to $n_s(0)$ exponentially) 
at low temperature.\cite{Landaeta-etal} The conventional $s$-wave pairing without any nodal excitation is naively expected from these experimental results.

We address this issue in this paper and show based on the multiband quasiclassical formalism \cite{Nagai-etal} that observed $T_1^{-1}$ and $n_s(T)$ are consistent with 
the chiral $d$-wave pairing as well as the $s$-wave one. 
The point is that the density of states (DOS) and root mean square of the Fermi velocity in the 3D band with nodal excitation are less dominant,\cite{Youn-etal,Youn-etal-2} and 
the power-law behavior in the low-temperature region is smeared out. 
It should also be emphasized that the HS peak is not only from the coherence effect solely exists for the conventional $s$-wave state, but also from the full gap structure of the quasiparticle excitation.\cite{Thinkham-textbook}  
Thus, an unconventional state without any nodes such as the chiral $d$-wave state in major quasi-2D bands is able to have a large HS peak. 
We also show that the measurement of the bulk quasiparticle DOS, which can be measured by the scanning tunneling spectroscopy/microscopy (STM/STS), 
would be crucial for the distinction between two pairing states. 

\begin{figure}[b]
\begin{center}
\centering
\includegraphics[width=\linewidth]{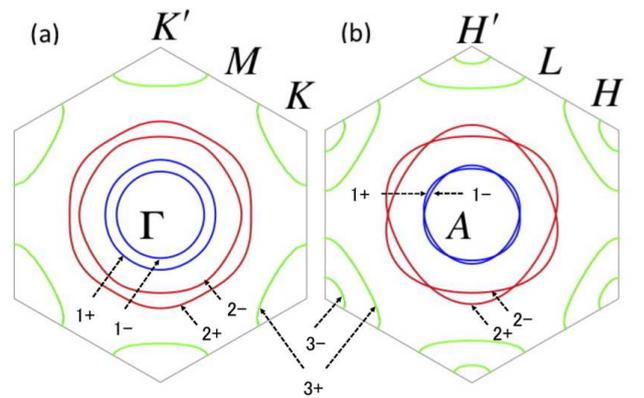}
\end{center} 
\caption{The cross section of Fermi surfaces at (a) $k_z=0$ and (b) $k_z=\pi / c$. 
Each Fermi surface is labelled by the set of parameters $\beta=1,2,3$ and $\gamma=\pm$ (see also Eq. (\ref{normal_spectrum})).}
\label{FS}
\end{figure}

\begin{table*}
\caption{The list of $\phi_{{\bm k}_F^{\beta\gamma}}$, which is the long-wavelength expansions (around the center of $``\beta\gamma"$-th Fermi surface) for $s$- and chiral $d$-wave pair wave functions 
in the tight-binding scheme.\cite{Goryo-Fischer-Sigrist, Fischer-Goryo} 
We omit the normalization constant from the condition $\langle |\phi_{{\bm k}_F}^{\beta\gamma}|^2 \rangle_{F^{\beta\gamma}}=1$, where $\langle \cdot\cdot\cdot \rangle_{F^{\beta\gamma}}$ denotes the average on the $``\beta\gamma"$-th Fermi surface.
The abbreviations $``3\pm(H)"$-th and $``3\pm(H')"$-th mean 
the disconnected Fermi pockets of ``$3\pm$"-th band enclosing $H$ and $H'$ points, respectively. Note that all the Fermi surfaces are quasi-2D, except for the 3D ``3-"-th one. 
Here, $\hat{\bm k}=\bm k/|\bm k|$, and 
$\delta{\bm k}=\bm k_F^{\beta\gamma}-\bm k_0$,
$\delta{\bm p}=\bm k_F^{\beta\gamma}-\bm p_0$,
$\delta{\bm p}'=\bm k_F^{\beta\gamma}-\bm p'_0$,
$\delta{\bm q}=\bm k_F^{\beta\gamma}-\bm q_0$,
and 
$\delta{\bm q}'=\bm k_F^{\beta\gamma}-\bm q'_0$ 
refer to the deviations from the centers of the long-wavelength expansions, 
and 
$\bm  k_0=(0,0,k_{F z}^{\beta\gamma})$,
$\bm  p_0=(2\pi/\sqrt{3},2\pi/3,k_{F z}^{\beta\gamma})$,
$\bm  p'_0=(0,4\pi/3,k_{F z}^{\beta\gamma})$,
$\bm  q_0=(2\pi/\sqrt{3},2\pi/3,\pi/c)$,
and 
$\bm  q'_0=(0,4\pi/3,\pi/c)$ the centers of the expansions. 
We emphasize that $\delta {\bm k}$, $\delta{\bm p}$, and $\delta{\bm p}'$ lie in the 2D plane, whereas $\delta{\bm q}$ and $\delta{\bm q}'$ point in 3D directions. 
Thus, $\phi_{{\bm k}_F^{3-}}$ for the chiral $d$-wave state has point nodes in the $k_z$-direction, while the others have no nodes. 
The linear dependence of the chiral $d$-wave function of $``3\pm"$-th bands is compatible with the Pauli exclusion principle due to the fact that 
we have an additional minus sign from the flipping of the valley degrees of freedom $H$ and $H'$.}
\begin{ruledtabular}
\begin{tabular}{c|cccccc}

& $``1\pm"$-th
& $``2\pm"$-th
& $``3+(H)"$-th
& $``3+(H')"$-th 
& $``3-(H)"$-th 
& $``3-(H')"$-th 
\\\hline
$\phi_{{\bm k}_F^{\beta\gamma}}$ of $s$-wave
& $1$
& $1$
& $1$
& $1$
& $1$
& $1$
\\
$\phi_{{\bm k}_F^{\beta\gamma}}$ of chiral $d$-wave
& $(\delta\hat{k}_{x} + i \delta\hat{k}_{y})^2$
& $(\delta\hat{k}_{x} + i \delta\hat{k}_{y})^2$
& $\delta\hat{p}_{x} - i \delta\hat{p}_{y}$
& $\delta \hat{p}'_{x} - i \delta\hat{p}'_{y}$
& $\delta\hat{q}_{x} - i \delta\hat{q}_{y}$
& $\delta \hat{q}'_{x} - i \delta\hat{q}'_{y}$

\end{tabular}
\end{ruledtabular}
\label{phi-k}
\end{table*}

\section{Normal and pairing states} 

There are two distinct honeycomb-shaped PtAs layers ($l=1,2$) in the unit cell of SrPtAs.\cite{Nishikubo-Kudo-Nohara}  
Although the entire crystal is inversion-symmetric, each layer does not contain the inversion center in itself, and the system is therefore staggered non-centrosymmetric.\cite{Sigrist-review} 
The band-structure calculation reveals that the Pt $5d$ orbital is dominant in the conduction bands and 
there are six Fermi surfaces with spin degeneracy, five of which are quasi-2D and the other 3D.\cite{Youn-etal,Youn-etal-2} 
Including Pt nearest-neighbor hopping within the plane, as well as nearest- and next-nearest-neighbor hopping 
between the planes, and also the staggered anti-symmetric SOC, 
one finds the one-body effective tight-binding Hamiltonian at low energy\cite{Youn-etal,Youn-etal-2}
\be
&&H_0=\sum_{\bm k \beta \sigma} 
\left(
\begin{array}{cc}
a_{\bm k 1 \sigma}^{(\beta)\dagger} & a_{\bm k 2 \sigma}^{(\beta)\dagger}
\end{array}
\right) 
\\
&&
 \ \ \ \ \ \ \ \ \ \ \times 
\left(
\begin{array}{cc}
\epsilon_{\bm k}^{(\beta)}+\alpha^{(\beta)}\lambda_{\bm k}\sigma & \epsilon_{c \bm k}^{(\beta)}
\\
 \epsilon_{c \bm k}^{(\beta)*} & \epsilon_{\bm k}^{(\beta)}-\alpha^{(\beta)}\lambda_{\bm k}\sigma
\end{array}
\right) 
\left(
\begin{array}{c}
a_{\bm k 1 \sigma}^{(\beta)}
\\
a_{\bm k 2 \sigma}^{(\beta)}
\end{array}
\right),
\nonumber
\ee
where $\beta(=1,2,3)$ indicates the unsplit band, $a_{\bm k l \sigma}^{(\beta)\dagger}$ ($a_{\bm k l \sigma}^{(\beta)}$) is the creation (annihilation) operator of an electron with the wave vector $\bm k$ and spin $\sigma=\pm 1$ in the $l$-th layer of the unit cell, 
$\epsilon^{(\beta)}_{1 \bm k}=t^{(\beta)}_{1} \sum_n \cos {\bm k}\cdot {\bm T}_n+t^{(\beta)}_{c2}\cos (k_z c)$, $\epsilon^{(\beta)}_{c \bm k}=t^{(\beta)}_{c} \cos (k_z c/2)[1+\exp(-i\bm k \cdot \bm T_3)+\exp(i\bm k \cdot \bm T_2)]$, and $\lambda_{\bm k}=\sum_n \sin \bm k \cdot {\bm T}_n$ 
with $\bm T_1=(0,a,0)$, $\bm T_2=(\sqrt{3}a/2,-a/2,0)$, and $\bm T_3=(-\sqrt{3}a/2,-a/2,0)$ the 
in-plane nearest-neighbor bond vectors ($a$ and $c$ are in-plane and inter-layer lattice constants).
Employing the unitary transformation $a^{(\beta)}_{\bm k l \sigma}=\sum U_{l\gamma} c^{(\beta)}_{\bm k \gamma \sigma}$ with band splitting 
$\gamma=\pm$, we diagonalize
$H_0=\sum \xi^{\beta \gamma}_{\bm k} c^{(\beta)\dagger}_{\bm k \gamma \sigma} c^{(\beta)}_{\bm k \gamma \sigma}$, where
\begin{eqnarray}
\xi^{\beta \gamma}_{\bm k}&=&\epsilon^{(\beta)}_{1\bm k}-\mu^{(\beta)} + \gamma \sqrt{|\epsilon^{(\beta)}_{c \bm k}|^2+|\alpha^{(\beta)} \lambda_{\bm k}|^2} 
\label{normal_spectrum}
\end{eqnarray}
is the normal-state energy spectrum. Using the tight-binding parameters suggested by the LDA calculation,\cite{Youn-etal,Youn-etal-2} we obtain the Fermi surface structure depicted as Fig. \ref{FS}.

We utilize the quasiclassical formalism for the multiband superconductor.\cite{Nagai-etal} Solving the Eilenberger equation, we obtain the quasiclassical Green's functions of the $``\beta\gamma"$-th bands
with the fermionic Matsubara energy $\epsilon_n=(2n+1) \pi k_B T$
\be
g_{\uparrow\uparrow}(i \epsilon_n, \bm k_F^{\beta\gamma})&=&\bar{g}_{\downarrow\downarrow}(i \epsilon_n, \bm k_F^{\beta\gamma})=\frac{\epsilon_n}{\sqrt{\epsilon_n^2+|\Delta_{\bm k_F^{\beta\gamma}}|^2}}, 
\\
f_{\uparrow\downarrow}(i \epsilon_n, \bm k_F^{\beta\gamma})&=&\left\{\bar{f}_{\downarrow\uparrow}(i \epsilon_n, \bm k_F^{\beta\gamma})\right\}^*=
\frac{\Delta_{\bm k_F^{\beta\gamma}}}{\sqrt{\epsilon_n^2+|\Delta_{\bm k_F^{\beta\gamma}}}|^2},   
\nonumber
\ee
where $\bm k_F^{\beta\gamma}$ is the Fermi wave vector and $\Delta_{\bm k_F^{\beta\gamma}}$ the gap function. We simply 
assume \cite{Nagai-etal,note-eq4}
\be
\Delta_{\bm k_F^{\beta\gamma}}&=&\Delta(T) \phi_{\bm k_F^{\beta\gamma}}, 
\label{gap}
\ee
where
\be
\Delta(T)=
\left\{
\begin{array}{cc}
\Delta_0 \tanh \left[\frac{\pi k_B T_c}{\Delta_0}\sqrt{\delta\left(\frac{T_c}{T}-1\right)}\right] & (T\leq T_c)
\\
0 & (T>T_c)
\end{array}
\right.,
\nonumber
\ee
with $\delta=1.05$, and $\phi_{\bm k_F^{\beta\gamma}}$ shown in Table \ref{phi-k} is the long-wavelength expansion (around the center of $``\beta\gamma"$-th Fermi surface)  
of the tight-binding pair wave functions for $s$- and chiral $d$-wave states, $1$ and $\sum_{n=1}^3 e^{i 2 \pi n /3} \cos \bm k \cdot \bm T_n$, respectively. \cite{Goryo-Fischer-Sigrist,Fischer-Goryo} 
We see from Table \ref{phi-k} that the chiral $d$-wave gap function in quasi-2D bands (the 3D band) has no nodes (point nodes in the $k_z$-direction).

We may introduce phenomenologically the quasiparticle damping (the smearing factor of the quasiparticle DOS) $\eta$ 
via the analytic continuation to obtain the retarded and advanced Green's functions 
\be
g_{\uparrow\uparrow}^{R,A}(\epsilon, \bm k_F^{\beta\gamma})&=&g_{\uparrow\uparrow}(i\epsilon_n \rightarrow \epsilon \pm i \eta, \bm k_F^{\beta\gamma}),
\label{Green-RA}\\ 
\bar{g}_{\downarrow\downarrow}^{R,A}(\epsilon, \bm k_F^{\beta\gamma})&=&\bar{g}_{\downarrow\downarrow}(i\epsilon_n \rightarrow \epsilon \pm i \eta, \bm k_F^{\beta\gamma}),
\nonumber\\
f_{\uparrow\downarrow}^{R,A}(\epsilon, \bm k_F^{\beta\gamma})&=&f_{\uparrow\downarrow}(i\epsilon_n \rightarrow \epsilon \pm i \eta, \bm k_F^{\beta\gamma}), 
\nonumber\\
\bar{f}_{\downarrow\uparrow}^{R,A}(\epsilon, \bm k_F^{\beta\gamma})&=&\bar{f}_{\downarrow\uparrow}(i\epsilon_n \rightarrow \epsilon \pm i \eta, \bm k_F^{\beta\gamma}). 
\nonumber
\ee
For simplicity, we neglect the band dependence of $\eta$. 
We therefore have two fitting parameters, $\Delta_0/k_B T_c$ and $\eta$, in the following calculations.

\begin{figure}
\begin{center}
\centering
\includegraphics[width=\linewidth]{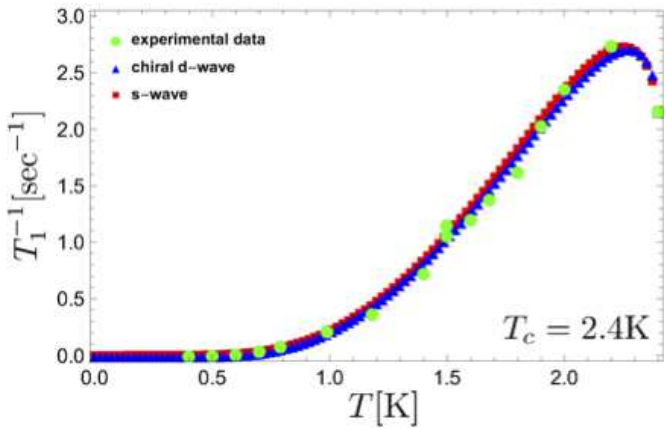}
\end{center} 
\caption{Temperature dependence of $T_1^{-1}$. Green dots are the experimental results.\cite{Matano-etal} Red squares and blue triangles show the estimations for $s$-wave and chiral $d$-wave states. 
Used fitting parameters are $\Delta_0/k_B T_c=1.765$ for both states, and $\eta=0.14 (0.008) k_B T_c$ for the $s$-wave (chiral $d$-wave) state. }
\label{T1}
\end{figure}

\begin{figure}
\begin{center}
\centering
\includegraphics[width=\linewidth]{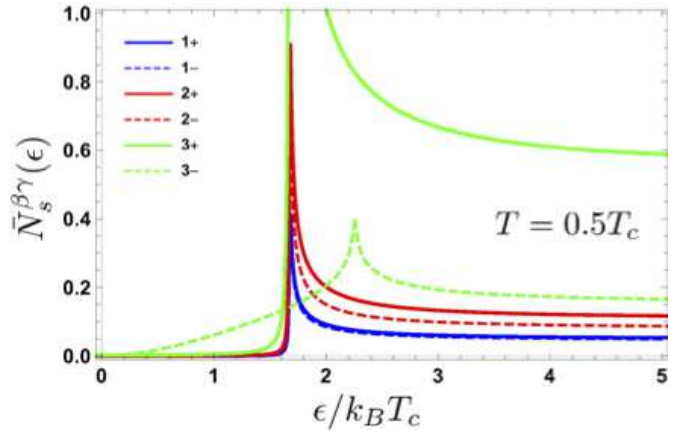}
\end{center} 
\caption{The contrubution from the ``$\beta\gamma$"-th band to the normalized quasiparticle DOS $\bar{N}_s(\epsilon)$ in the chiral $d$-wave state, which is expressed as 
$\bar{N}_s^{\beta\gamma}(\epsilon)= \int a_{\uparrow\uparrow}^{11}(\epsilon, \bm k_F^{\beta\gamma}) d \Omega_{\bm k_{F}}^{\beta\gamma} /\left\{(2 \pi)^3 N(0) \hbar |\bm v_{F}^{\beta\gamma}|\right\}$. 
The gap amplitude at $T=0.5T_c$ is used in this estimation.  
We see that the 3D band (``3-"-th band) is less dominant. 
We note additionally that the gap maxima, which determines the peak location, of the point nodal gap function $\Delta(T) \phi_{{\bm k}_F^{3-}}$ is larger than 
that of the other nodeless gap functions due to the normalization condition for the Fermi surface average 
$\langle |\phi_{{\bm k}_F}^{\beta\gamma}|^2 \rangle_{F^{\beta\gamma}}=1$ (see also Eq. \ref{gap}).}
\label{SC_DOS-each}
\end{figure}

\section{The nuclear-spin-lattice relaxation rate $T_1^{-1}$} 

The relaxation rate is \cite{Nagai-etal} 
\be
\frac{T_1(T_c)}{T_1(T)}&=&\frac{T}{T_c} \int_{-\infty}^{\infty} d \epsilon \left(\bar{N}_{s}(\epsilon)^2+\bar{M}_{s}(\epsilon)^2\right)\left(-\frac{\partial f(\epsilon)}{\partial \epsilon}\right), 
\label{1/T1}
\ee
where $f(\epsilon)$ is the Fermi-Dirac distribution function, and 
$\bar{N}_s(\epsilon)$ and $\bar{M}_s(\epsilon)$ denote DOS and anomalous DOS of the Bogoliubov quasiparticle normalized by the entire DOS at the Fermi level in the normal state $N(0)$. 
For the multi-band spin-singlet superconductor, \cite{Nagai-etal}
\be
\left\{
\begin{array}{l}
\bar{N}^2_{s}(\epsilon)=\langle a_{\uparrow\uparrow}^{11}(-\epsilon, \bm k) \rangle_F \langle a_{\downarrow\downarrow}^{22}(\epsilon, \bm k) \rangle_F, 
\\
\bar{M}^2_{s}(\epsilon)=-\langle a_{\uparrow\downarrow}^{12}(-\epsilon, \bm k) \rangle_F \langle a_{\downarrow\uparrow}^{21}(\epsilon, \bm k) \rangle_F,
\end{array}
\right.
\ee
where 
\be
a_{\uparrow\uparrow}^{11}(\epsilon, \bm k)&=&\frac{1}{2} \left(g_{\uparrow\uparrow}^R(\epsilon, \bm k)-g_{\uparrow\uparrow}^A(\epsilon, \bm k)\right), 
\\ 
a_{\downarrow\downarrow}^{22}(\epsilon, \bm k)&=&\frac{1}{2}\left(\bar{g}_{\downarrow\downarrow}^R(\epsilon, \bm k)-\bar{g}_{\downarrow\downarrow}^A(\epsilon, \bm k)\right),
\nonumber\\  
a_{\uparrow\downarrow}^{12}(\epsilon, \bm k)&=&\frac{i}{2} \left(f_{\uparrow\downarrow}^R(\epsilon, \bm k)- f_{\uparrow\downarrow}^A(\epsilon, \bm k)\right), 
\nonumber\\
a_{\downarrow\uparrow}^{21}(\epsilon, \bm k)&=&\frac{i}{2} \left(\bar{f}_{\downarrow\uparrow}^R(\epsilon, \bm k)- \bar{f}_{\downarrow\uparrow}^A(\epsilon, \bm k)\right),  
\nonumber
\ee 
and
\be
\langle a_{\sigma\sigma'}^{\tau\tau'}(\epsilon, \bm k) \rangle_F&=& \frac{1}{N(0)}\sum_{\beta\gamma} \int \frac{d \Omega_{\bm k_{F}^{\beta\gamma}}}{(2 \pi)^3 \hbar |\bm v_{F}^{\beta\gamma}|} a_{\sigma\sigma'}^{\tau\tau'}(\epsilon, \bm k_F^{\beta\gamma}) 
\nonumber\\
&&\hbar \bm v_F^{\beta\gamma}=\left.{\bm \nabla}_{\bm k} \xi_{\bm k}^{\beta\gamma}\right|_{\bm k=\bm k_F^{\beta\gamma}} 
\ee
is the Fermi surface average.

The results for both $s$-wave and chiral $d$-wave states are shown in Fig. \ref{T1} with experimental data.\cite{Matano-etal}
The fitting parameters are chosen as $\Delta_0/k_B T_c=1.765$ for both states, and $\eta=0.14 (0.008) k_B T_c$ for the $s$-wave (chiral $d$-wave) state.
We clearly see that both pairing states agree well with experimental data showing the HS peak just below $T_c$ and 
exponential decay at low temperature.

It is renown that $\bar{M}_s(\epsilon)$ from the coherence effect appears only for the $s$-wave state and contributes to the HS peak significantly.\cite{Thinkham-textbook} 
We should note, however, that the quasiparticle excitations of quasi-2D bands in the chiral $d$-wave state is fully gapped and $\bar{N}_s(\epsilon)$ also gives rise to the
large enough HS peak with reduced $\eta$.
Moreover, the quasiparticle DOS of the 3D band is less dominant in this system (see Fig. \ref{SC_DOS-each}), and then the power-low behavior 
at low temperature caused by the nodal excitation is negligible. These facts are crucial for the compatibility of the chiral $d$-wave state with experiment data.

\begin{figure}
\begin{center}
\centering
\includegraphics[width=\linewidth]{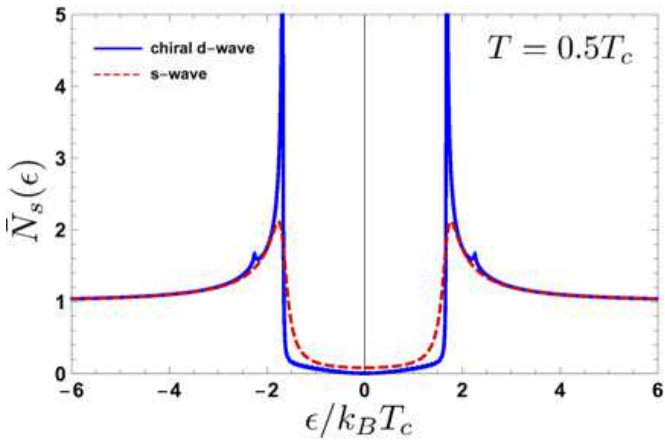}
\end{center} 
\caption{Dashed red and blue lines are the normalized DOS of quasiparticles $\bar{N}_s(\epsilon)$ in $s$- and chiral $d$-wave states with the gap amplitude at $T=0.5T_c$. 
The smearing factor $\eta=0.14 (0.008) k_B T_c$, and the peaks are reduced (enhanced) in the $s$-wave (chiral $d$-wave) state. 
The large difference between the magnitudes of the peaks would be crucial for the experimental distinction of these two pairing states.
Incidentally, the point-nodal excitation of the chiral $d$-wave state from the less dominant 3D band causes 
feeble ``V-shaped" behavior around $\epsilon=0$ and tiny peaks at $\epsilon \simeq \pm 2.3 k_B T_c$ in the blue line.}
\label{SC_DOS}
\end{figure}

We therefore cannot distinguish between $s$- and chiral $d$-wave states from $T_1^{-1}$. 
{\it We then propose that the measurement of the bulk quasiparticle DOS would give a decisive distinction.} 
It should be emphasized that we need to reduce $\eta$ for the chiral $d$-wave state to compensate the absence of the contribution from $\bar{M}_s(\epsilon)$. Namely,   
the reduction of $\eta$ causes the significant difference of $\bar{N}_s(\epsilon)$ for $s$- and chiral $d$-wave states (see Fig. \ref{SC_DOS}). 
Thus, the observation of $\bar{N}_s(\epsilon)$ 
would be relevant for the distinction between two pairing states using, for instance, the 
STM/STS even in the (0001) surface without the chiral surface mode. 
It would be an advantage, since the layered structure of the crystal enables us to obtain a flat (0001) surface. 
The other surfaces may have more roughness and be not suitable for the measurement, although 
they host the chiral surface states causing zero-energy peak as the fingerprint of the topological chiral $d$-wave pairing. 
We note that another possibility for the distinction using quasiparticle interference spectroscopy has also been pointed out.\cite{Akbari-Thalmeier}

\begin{figure}
\begin{center}
\centering
\includegraphics[width=\linewidth]{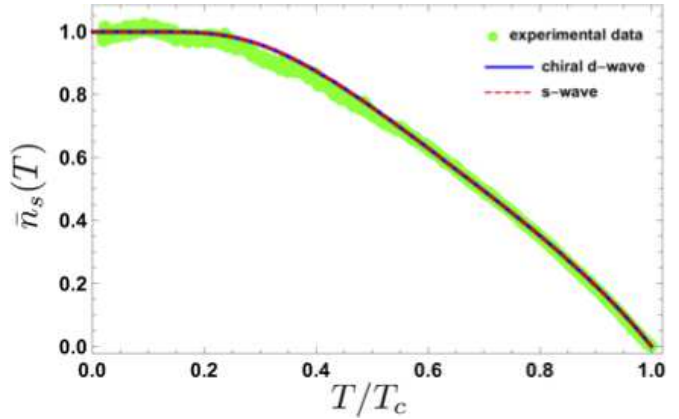}
\end{center} 
\caption{Temperature dependence of normalized superfluid density $\bar{n}_s(T)$. Green dots denote experimental data.\cite{Landaeta-etal} 
Dashed red and blue lines show the estimations for $s$-wave and chiral $d$-wave states. 
Used fitting parameters are $\Delta_0/k_B T_c=1.5$ for both states, and $\eta=0.14 (0.008) k_B T_c$ for the $s$-wave (chiral $d$-wave) state. 
We note additionally that the results are insensitive to the choice of $\eta$ in this case.}
\label{n_s}
\end{figure}

\begin{figure}
\begin{center}
\centering
\includegraphics[width=\linewidth]{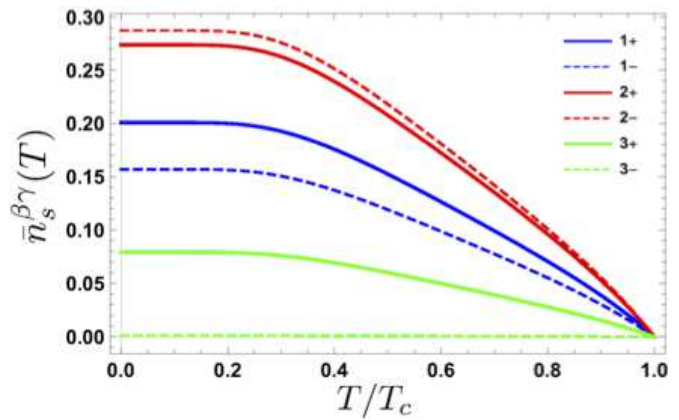}
\end{center} 
\caption{$\bar{n}_s^{\beta\gamma}(T)$ of the chiral $d$-wave state. We see that the contribution from the 3D band (``3-"-th band) with power-law behavior is negligibly small.}
\label{FS-dep-n_s}
\end{figure}

\begin{table*}
\caption{The list of $\phi_{{\bm k}_F^{\beta\gamma}}$ for the $f$-wave state. Here, $\hat{\bm k}=\bm k/|\bm k|$, $\delta{\bm k}=\bm k_F^{\beta\gamma}-\bm k_0$, and 
$\bm  k_0=(0,0,k_{F z}^{\beta\gamma})$.}
\begin{ruledtabular}
\begin{tabular}{c|cccccc}

& $``1\pm"$-th
& $``2\pm"$-th
& $``3+(H)"$-th
& $``3+(H')"$-th 
& $``3-(H)"$-th 
& $``3-(H')"$-th 
\\\hline
$\phi_{{\bm k}_F^{\beta\gamma}}$ of $f$-wave
& $(3 \delta \hat{k}_x^2-\delta \hat{k}_y^2) \delta \hat{k}_y$
& $(3 \delta \hat{k}_x^2-\delta \hat{k}_y^2) \delta \hat{k}_y$
& $1$
& $-1$
& $1$
& $-1$
\end{tabular}
\end{ruledtabular}
\label{phi-k-f}
\end{table*}

\begin{figure*}
\begin{minipage}{0.47\hsize}
\begin{center}
\centering
\includegraphics[width=\linewidth]{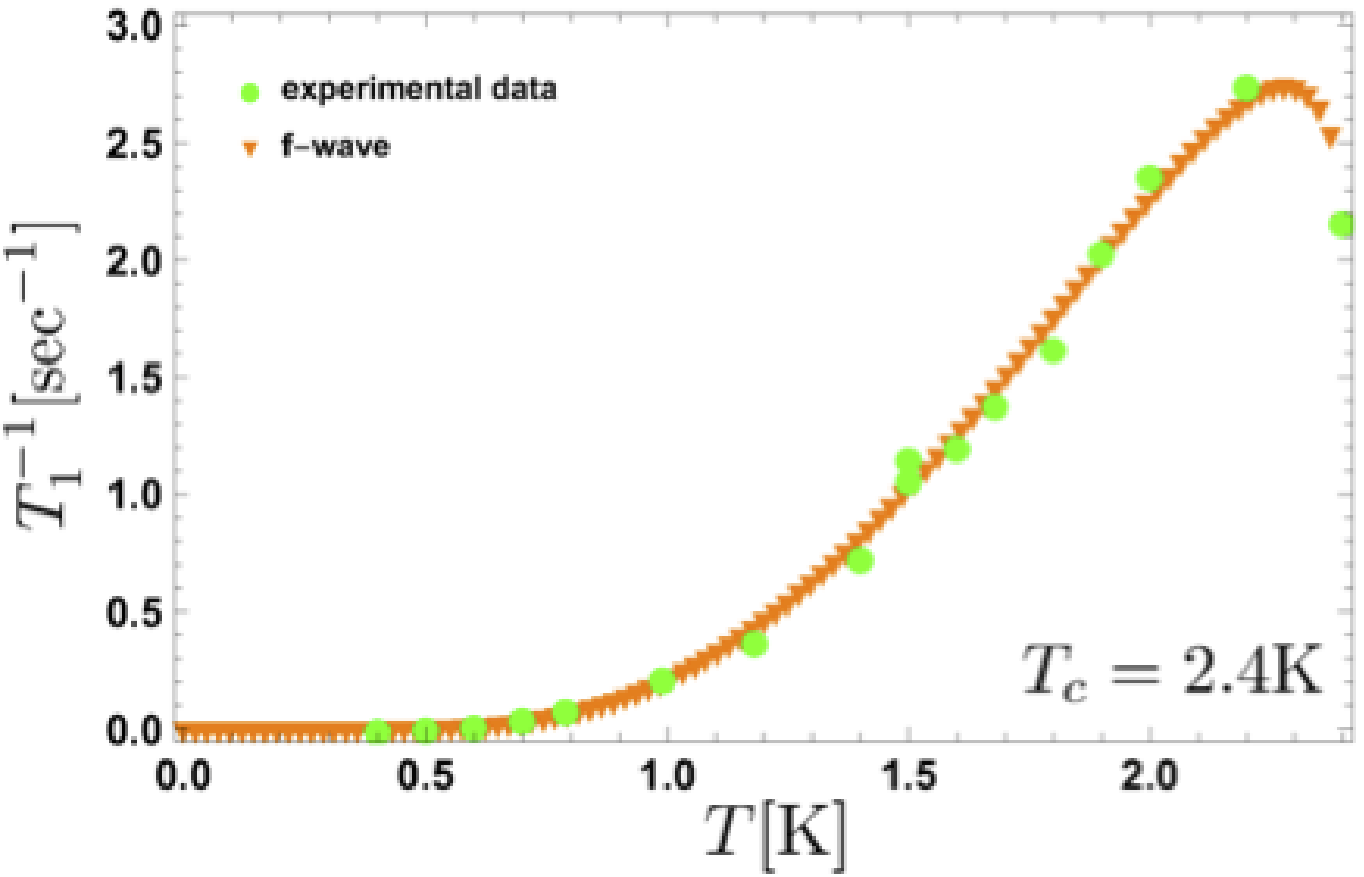}
\end{center} 
\caption{Temperature dependence of $T_1^{-1}$. Green dots are the experimental results,\cite{Matano-etal} and orange triangles show the estimation for the $f$-wave state. 
Used fitting parameters are $\Delta_0/k_B T_c=1.765$ and $\eta=0.0025k_B T_c$. }
\label{T1-f}
\end{minipage}
\ \ \ \ \ \ \ \ 
\begin{minipage}{0.47\hsize}
\begin{center}
\centering
\includegraphics[width=\linewidth]{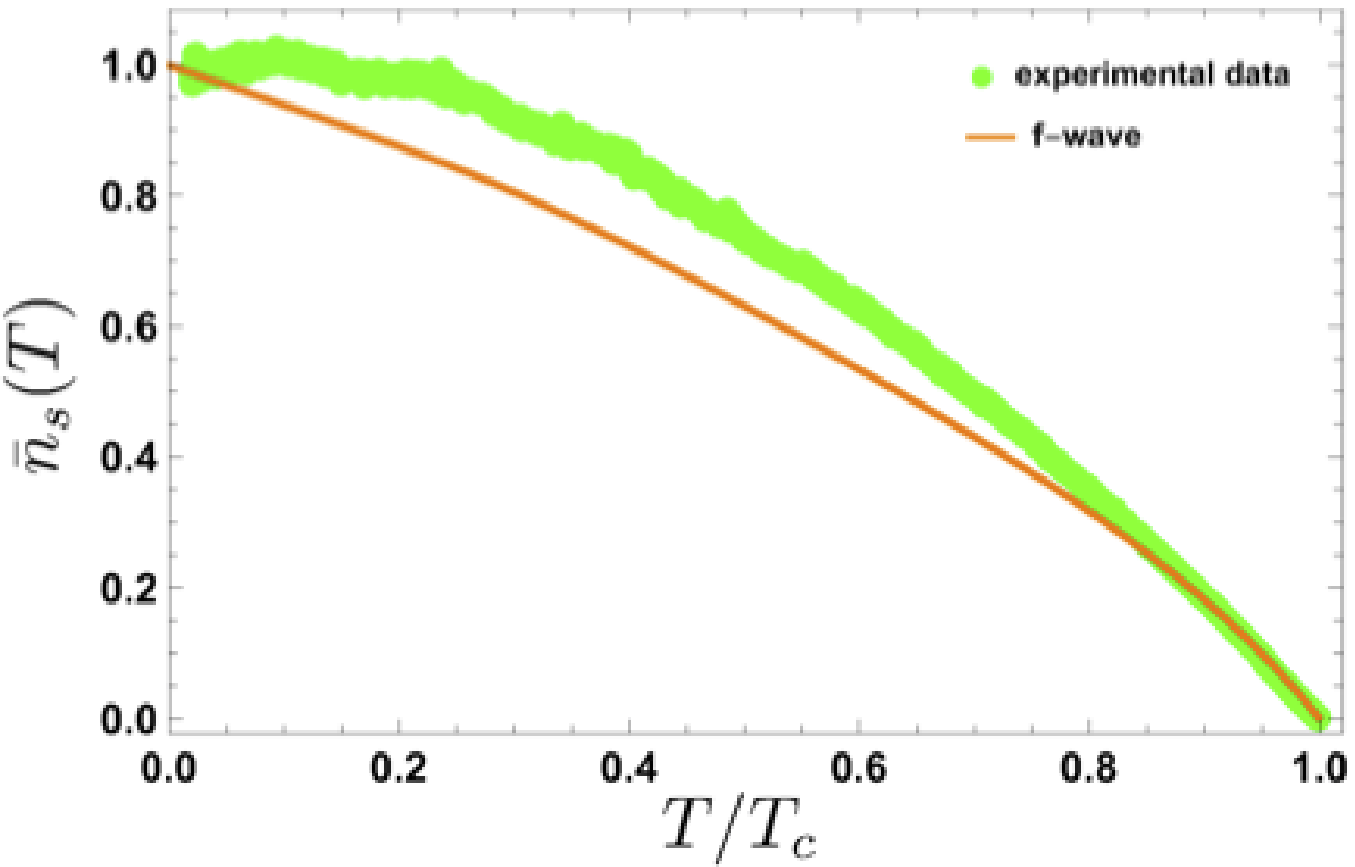}
\end{center} 
\caption{Temperature dependence of normalized superfluid density $\bar{n}_s(T)$. Green dots denote experimental data,\cite{Landaeta-etal} 
and the orange line shows the estimation for the $f$-wave state. 
Used fitting parameters are $\Delta_0/k_B T_c=1.5$ and $\eta=0.0025k_B T_c$.}
\label{n_s-f}
\end{minipage}
\end{figure*}

\section{Superfluid density $n_s (T)$}
 
Superfluid density normalized by its zero-temperature value $n_{s}(0)$ is\cite{Nagai-etal} 
\be
\bar{n}_s(T)&=&\sum_{\beta\gamma} \bar{n}_s^{\beta\gamma}(T), 
\label{sup-dens}
\\
\bar{n}_s^{\beta\gamma}(T)&=&\frac
{\sum_{i=x,y,z} \int \frac{d \Omega_{\bm k_{F}^{\beta\gamma}}}{(2 \pi)^3 \hbar |\bm v_{F}^{\beta\gamma}|}   \left(v_{Fi}^{\beta\gamma}\right)^2 \left(1 - Y_{\bm k_F^{\beta\gamma}}(T)\right)}
{\sum_{\beta\gamma} \sum_{i=x,y,z} \int \frac{d \Omega_{\bm k_{F}^{\beta\gamma}}}{(2 \pi)^3 \hbar |\bm v_{F}^{\beta\gamma}|}   \left(v_{Fi}^{\beta\gamma}\right)^2}, 
\nonumber
\ee
where
\be
Y_{\bm k}(T)=1-\pi k_B T \sum_{n=-\infty}^{\infty} \frac{|\Delta_{\bm k}|^2}{\left(\epsilon_n^2 + |\Delta_{\bm k}|^2\right)^{2/3}}
\nonumber
\ee
is Yosida function. The parameter is taken as $\Delta_0/k_B T_c=1.5$ for both states.\cite{note} 
We have also checked that the results are insensitive to the choice of the quasiparticle damping $\eta$ in this case, and we take the values $\eta=0.14 (0.008) k_B T_c$ for the
$s$-wave (chiral $d$-wave) state, the same ones used in the estimations of $T_1^{-1}$.  

We clearly see from Fig. \ref{n_s} that the results of both states fit very well to experimental data, namely, both exhibit the thermal-activation-type behavior at low temperature. 
$\bar{n}_s^{\beta\gamma}(T)$ in Eq. (\ref{sup-dens}) shows the contribution from each band and the result for the chiral $d$-wave state is plotted in Fig. \ref{FS-dep-n_s}. 
The contribution from the 3D band with power-law behavior is negligibly small, since $\bar{n}_s^{\beta\gamma}(T)$
depends strongly on root mean square of the Fermi velocity, and its value for the 3D band is minor (see Table. I in Ref. \onlinecite{Youn-etal-2}).

\section{Summary}
 
We have shown based on the multiband quasiclassical formalism\cite{Nagai-etal} that observed $T_1^{-1}$ and $n_s(T)$ in the superconducting phase of SrPtAs\cite{Matano-etal,Landaeta-etal} 
are consistent with the chiral $d$-wave pairing as well as the $s$-wave one. 
In other words, the chiral $d$-wave state cannot be ruled out from these experiments. 
We have found in the fitting of $T_1^{-1}$ a significant difference of 
the quasiparticle damping factors for these two pairing states due to the absence of $\bar{M}_s(\epsilon)$ in the chiral $d$-wave state (see Eq. (\ref{1/T1})). 
This difference causes a remarkable difference between the magnitudes of the peaks in the bulk quasiparticle DOS, therefore, a measurement of which would 
give a decisive distinction between $s$- and chiral $d$-wave states (see Fig. \ref{SC_DOS}). 
Such a measurement could be done by the STM/STS even in the (0001) surface without the chiral surface mode. 
It would be an advantage, since we may obtain a flat (0001) surface due to the layered structure of the crystal. 
The essence of our results comes from the fact that the DOS and root mean square of Fermi velocity 
are less dominant in the 3D ($``3-"$-th) band,\cite{Youn-etal,Youn-etal-2} and  
would be robust even if we take a further approximation with respect to parameters in the superconducting state.

We comment on the $f$-wave pairing suggested as the other possibility.\cite{f-wave}
The quasiparticle excitation of this state is fully gapped in two bands ($``3\pm"$-th) around the Brillouin zone corners, whereas has line nodes in four quasi-2D bands ($``1\pm"$-th and $``2\pm"$-th)
around the zone center. We have checked that $T_1^{-1}$ for the $f$-wave state using the smallest $\eta=0.0025k_B T_c$ fits well with observed data,\cite{Matano-etal}
thanks to the large DOS of fully-gapped $``3+"$-th band.  
However, $n_s(T)$, the dominant contribution for which comes from four line-nodal $``1\pm"$-th and $``2\pm"$-th bands with large 
root mean square of the Fermi velocity (see Table. I in Ref. \onlinecite{Youn-etal-2}),  
shows an evident power-law behavior at low temperature and contradicts strongly with the experiment.\cite{Landaeta-etal} The results are summarized in the supplement. \cite{supp}  
Besides, it would be hard to explain the spontaneous magnetization\cite{Biswas-etal} from the $f$-wave state as well as the $s$-wave one. 
The chiral $d$-wave state is thus the only one, which is consistent with all the experiments that have been done so far.\cite{Biswas-etal,Matano-etal,Landaeta-etal} 

We should also mention that only poly-crystal samples have ever been used. 
The experiments with single crystals are highly desired.

\acknowledgments 
The authors thank M. Nohara, K. Kudo, and M. Sigrist for their useful discussions, and I. Bonalde and K. Matano for sending their experimental data. 
This work was partially supported by JSPS KAKENHI Grant No. 15H05885 (J-Physics).  
J.G. is grateful to the Pauli Center for Theoretical Physics of ETH Zurich for hospitality.

\appendix
\begin{center}
{\bf{Appendix: $T_1^{-1}$ and $n_s(T)$ for the $f$-wave state}}
\end{center}

We comment on the $f$-wave state\cite{f-wave} suggested as the other possibility of the pairing symmetry in the hexagonal pnictide superconductor SrPtAs.\cite{Nishikubo-Kudo-Nohara}
The quasiparticle excitation of this state is fully gapped in two bands ($``3\pm"$-th) around the Brillouin zone corners, whereas has line nodes in four quasi-2D bands ($``1\pm"$-th and $``2\pm"$-th) 
around the zone center. The function $\phi_{\bm k_F^{\beta\gamma}}$ for the $f$-wave state is listed in Table \ref{phi-k-f} in this Appendix. 

We show, in Fig. \ref{T1-f}, $T_1^{-1}$ for the $f$-wave state using the parameters $\Delta_0/k_B T_c=1.765$ and $\eta=0.0025k_B T_c$. 
We see the result fits well with observed data,\cite{Matano-etal} thanks to the large DOS of fully-gapped $``3+"$-th band.  
However, $n_s(T)$ in Fig. \ref{n_s-f} shows an evident power-law behavior at low temperature and contradicts strongly with the experiment.\cite{Landaeta-etal} 
The power-law behavior comes from the fact that line-nodal $``1\pm"$-th and $``2\pm"$-th bands with large 
root mean square of the Fermi velocity 
(see Table. I in Ref. \onlinecite{Youn-etal-2}) gives the dominant contribution to $n_s(T)$.


\begin{thebibliography}{99}
\bibitem{note-eq4} 
Neglecting the band dependence of the gap amplitude $\Delta(T)$ is actually an oversimplification, but it would not be crucial. 
We would obtain essentially the same results by introducing the band dependence of the quasiparticle damping (see $\eta$ in Eq. (\ref{Green-RA})), even if
we take band-dependent gap amplitudes into account.
\bibitem{note} We have taken different values of the fitting parameter $\Delta_0/k_B T_c$ for $T_1^{-1}$ and $n_s(T)$. 
It seems to be natural, since different samples are used in these experiments.\cite{Matano-etal,Landaeta-etal}
\end{thebibliography}
\end{document}